\documentclass[aps, prl, twocolumn, superscriptaddress, showpacs]{revtex4}
\usepackage{amsfonts}
\usepackage{bbm}

\usepackage{graphicx}
\begin{document}

\title{Spin-Selective Transport of Electron in DNA Double Helix}

\author{Ai-Min Guo}
\affiliation{Institute of Physics, Chinese Academy of Sciences,
Beijing 100190, China}
\author{Qing-feng Sun}
\email{sunqf@iphy.ac.cn}
\affiliation{Institute of Physics, Chinese
Academy of Sciences, Beijing 100190, China}

\date{\today}

\begin{abstract}
The experiment that the high spin selectivity and the
length-dependent spin polarization are observed in double-stranded
DNA [Science ${\bf 331}$, 894 (2011)], is elucidated by considering
the combination of the spin-orbit coupling, the environment-induced
dephasing, and the helical symmetry. We show that the spin
polarization in double-stranded DNA is significant even in the case
of weak spin-orbit coupling, while no spin polarization appears in
single-stranded DNA. Furthermore, the underlying physical mechanism
and the parameters-dependence of the spin polarization are studied.
\end{abstract}


%Use showkeys class option if keyword
%display desired
\maketitle

Molecular spintronics, by combining molecular electronics with
spintronics to manipulate the transport of electron spins in organic
molecular systems, is regarded as one of the most promising research
fields and is now attracting extensive interest
\cite{rar,fa,dva,um}, owing to the long spin relaxation time and the
flexibility of organic materials. Unconventional magnetic properties
of molecular systems reported in organic spin valves and magnetic
tunnel junctions, are attributed to the hybrid states in the
organic-magnetic interfaces \cite{xzh,bc,bj,an,ss} and to
single-molecule magnet \cite{um}. Organic molecules would not be
suitable candidates for spin-selective transport because of their
nonmagnetic properties and weak spin-orbit coupling (SOC) \cite{kf}.

However, very recently, G\"{o}hler {\it et al.} reported the spin
selectivity of photoelectron transmission through self-assembled
monolayers of double-stranded DNA (dsDNA) deposited on gold
substrate \cite{gb}. They found that well-organized monolayers of
the dsDNA act as very efficient spin filters with high spin
polarization at room temperature for long dsDNA, irrespective of the
polarization of the incident light. The spin filtration efficiency
increases with increasing length of the dsDNA and contrarily no spin
polarization could be observed for single-stranded DNA (ssDNA).
These results were further substantiated by direct charge transport
measurements of single dsDNA connected between two leads \cite{xz}.
Although several theoretical models were put forward to investigate
the spin-selective properties of DNA molecule based on single
helical chain-induced Rashba SOC \cite{ys,gr}, the models neglect
the double helix feature of the dsDNA and are somewhat inconsistent
with the experimental results that the ssDNA could not be a spin
filter. Until now the underlying physical mechanism remains unclear
for high spin selectivity observed in the dsDNA \cite{rg,dvm}.

In this Letter, a model Hamiltonian, including the small
environment-induced dephasing, the weak SOC, and the helical
symmetry, is proposed to investigate the quantum spin transport
through the ssDNA and dsDNA connected to nonmagnetic leads. We
interpret the experimental results that the electrons transmitted
through the dsDNA exhibit high spin polarization, the spin
filtration efficiency will be enhanced by increasing the DNA length,
and no spin polarization appears for the ssDNA. The physical
mechanism arises from the combination of the dephasing, the SOC, and
the helical symmetry. No spin polarization could be observed if any
aforementioned factor is absent. In addition, the spin polarization
could be considerably enhanced by appropriately increasing the
dephasing strength or by decreasing the helix angle.

\begin{figure}
\includegraphics[scale=0.77]{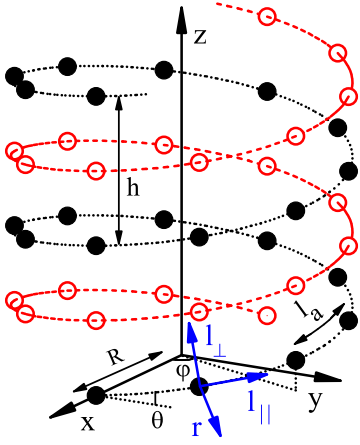}
\caption{\label{fig:one}(color online). Schematic view of the dsDNA
with radius $R$, pitch $h$, helix angle $\theta$, and arc length
$l_a$. The circles represent the nucleobases, where the full ones
assemble one helical chain and the open ones form the other helical
chain. The arc length satisfies $l_a\cos\theta = R \Delta\varphi$
and $l_a\sin\theta=\Delta h$, with $\Delta\varphi$ and $\Delta h$
being the twist angle and the stacking distance between neighboring
base pairs, respectively. We set $R=1$ nm, $\Delta h = 0.34$ nm, and
$\Delta\varphi= {\frac \pi 5}$, which are typical values of B-form
DNA. Other parameters can be obtained: $h = 3.4$ nm, $\theta \approx
0.5$ rad, and $l_a\approx0.71$ nm. }
\end{figure}

The charge transport through the dsDNA, illustrated in
Fig.~\ref{fig:one}, can be simulated by the Hamiltonian:
\begin{equation}
{\cal H}={\cal H}_{\rm DNA} +{\cal H}_{lead}+{\cal H}_{c} + {\cal
H}_{\rm so} +{\cal H}_d . \label{eq:one}
\end{equation}
Here $ {\cal H}_{\rm DNA} = \sum_{n=1}^N [\sum_{j=1} ^2 (
\varepsilon_ {jn} c_{jn}^\dag c_{jn} + t_{jn} c_{jn}^\dag c_{jn+1})
+\lambda_n c_{1n}^\dag c_{2n} +\mathrm{H.c.}] $ is the Hamiltonian
of usual two-leg ladder model including spin degree of freedom
\cite{cg}, with $N$ the DNA length, $c_{jn} ^\dag=
(c_{jn\uparrow}^\dag, c_{ jn \downarrow } ^\dag)$ the creation
operator of the spinor at the $n$th site of the $j$th chain of the
dsDNA, $\varepsilon_ {jn} $ the on-site energy, $t_{jn}$ the
intrachain hopping integral, and $\lambda_{n}$ the interchain
hybridization interaction. ${\cal H}_{lead}+{\cal H}_{c} =
\sum_{k,\beta (\beta={\rm L}, {\rm R})} [ \varepsilon_{\beta k}
a_{\beta k}^\dag a_{\beta k} + t_{\beta} a_{\beta k}^\dag (c_{1
n_{\beta}} +c_{2 n_{\beta}})+ \mathrm {H.c.} ]$ describe the left
and right nonmagnetic leads and the coupling between the leads and
the dsDNA, with $n_{\rm L}=1$ and $n_{\rm R}=N$. ${\cal H}_{\rm so}$
and ${\cal H}_d$ are, respectively, the Hamiltonians of the SOC term
and the dephasing term, which will be discussed in the following.

When a charge is moving under an electrostatic potential $V$, an SOC
arises ${\cal H}_{\rm so}={\frac \hbar {4m^2c^2} }\nabla V \cdot(
\hat {\sigma} \times \hat{\vec{p}} )$, with the electron mass $m$,
the speed of light $c$, the Pauli matrices $\hat {\sigma} = ({\bf
\sigma}_x, {\bf \sigma}_y, {\bf \sigma}_z)$, and the momentum
operator $\hat{\vec{p}}$. In the dsDNA, the differences of the
potential are usually bigger along the radial direction $\hat{r}$
than that along the helix axis ($z$-axis in Fig.~\ref{fig:one})
\cite{hd}. On the other hand, since the differences of $V$ are
especially large between the interior and the exterior of the dsDNA,
$dV/dr$ is very large at the boundary $r=R$ with $R$ the radius
\cite{note1}. Hence, it is reasonable to consider the $r$-component
of $V$ only and the SOC can be simplified in the cylindrical
coordinate system ${\cal H}_{\rm so}=-{\frac \alpha \hbar} \hat{
\sigma } \cdot ( \hat{r} \times \hat{\vec{p}} )$ with $\alpha \equiv
{\frac {\hbar^2} {4m^2c^2}} {\frac d {dr}}V(r)$. Considering a
charge propagating in one helical chain of the dsDNA (e.g., the
dotted line in Fig.~\ref{fig:one}), the momentum $\hat{ \vec{p} }=
\hat{p} _ \parallel {\hat l}_\parallel$ with ${\hat l} _\parallel $
the unit vector along the helical chain direction. Thus ${\cal H}_
{\rm so}$ is reduced to ${\cal H} _{\rm so} = - \frac {\alpha}
{2\hbar} [ \sigma _{\perp} \hat{p}_{\parallel} +\hat{p}_ {\parallel}
\sigma_ { \perp } ]$, where $\sigma_ {\perp} (\varphi) = \sigma_ x
\sin\varphi \sin \theta -\sigma_ y \cos\varphi \sin\theta +\sigma_z
\cos\theta$ with $\theta$ the helix angle and $\varphi$ the
cylindrical coordinate. Since the dsDNA consists of two helical
chains, the total SOC is ${\cal H}_ {\rm so} =-\frac{\alpha}{2\hbar}
\sum_ {j=1 } ^ 2 [\sigma_ {\perp}^ {(j)} \hat{p}_ {\parallel}^{(j)}
+\hat {p} _{ \parallel}^{(j)} \sigma_ {\perp}^ {(j)} ]$ with
$\sigma_ {\perp}^ {(1)}= \sigma_ {\perp} (\varphi)$ and $\sigma_
{\perp} ^{(2)} = \sigma _{\perp} (\varphi +\pi)$. The interchain SOC
has been neglected because it is very small due to the potential
symmetry. By using the second quantization \cite{sqf1}, ${\cal
H}_{\rm so}$ can be written as:
\begin{equation}
{\cal H}_{\rm so}= \sum_{j,n} i t_{\rm so} c_{jn}^\dag [ \sigma _n ^
{(j)}+ \sigma_{n+1}^{(j)}] c_{jn+1}+ \mathrm{H.c.} ,\label{eq:two}
\end{equation}
where $t_{\rm so}=-{\frac \alpha {4 l_{a} }}$, $\sigma_ {n+1} ^{(1)
}= \sigma_{\perp}(n\Delta\varphi)$, and $ \sigma_ {n+1}^ {(2)} =
\sigma _ {\perp}(n\Delta\varphi +\pi)$. $l_a$ and $\Delta \varphi$
are, respectively, the arc length and the twist angle between
successive base pairs.

On the other hand, a charge transmitting through the dsDNA will
experience inelastic scattering from the phonons due to the
fluctuation of each nucleobase around its equilibrium position and
other inelastic collisions with the absorbed counterions in the
dsDNA due to the negatively charged sugar-phosphate backbones
\cite{lxq}. Such inelastic scattering will give rise to the lose of
the phase and spin memory of the charge. To simulate the
phase-breaking process, B\"{u}ttiker's virtual lead is introduced by
connecting to each nucleobase \cite{xy,jh}, with the Hamiltonian of
the dephasing term being:
\begin{equation}
{\cal H}_d = \sum_{j,n,k} ( \varepsilon_{jnk}a_{jnk}^\dag a_{jnk}  +
t_d a_{jnk}^\dag c_{jn} +\mathrm{H.c.}). \label{eq:three}
\end{equation}
$a_{jnk} ^\dag= (a_{jnk\uparrow} ^\dag, a_{jnk \downarrow } ^\dag)$
is the creation operator of the virtual lead and $t_d$ is the
coupling between the nucleobase and the virtual lead.

Let us demonstrate analytically that the ssDNA could not behave as a
spin filter. In continuous real-space spectrum, the Hamiltonian of
the ssDNA containing the SOC term is written as ${\cal H}_{\rm ss}=
{\frac {\hat{p}_{\parallel}^2} {2m}} -\frac{\alpha}{2\hbar}
[\sigma_{\perp} \hat{p}_{\parallel} +\hat{p}_{\parallel}
\sigma_{\perp} ] + V(l)$ with $V(l)$ the potential energy of the
helical chain. By taking a unitary transformation with the operator
$U(l)=e^{ (i m \alpha /\hbar^2) \int_l \sigma_{\perp} dl }$
\cite{sqf2}, ${\cal H}_{\rm ss}$ is transformed into ${\cal H}_{\rm
ss} ' = U^{\dag}H_{\rm ss} U = {\frac {\hat{p}_{\parallel}^2} {2m}}
- {\frac {m \alpha^2 } {2 \hbar^2}} + V(l)$, which is independent of
spin. Therefore, no spin polarization could be observed in the
ssDNA, regardless of the SOC term, the existence of the dephasing,
and other model parameters. This result can be obtained also by
using the discrete Hamiltonians of Eqs.~(\ref{eq:one})
and~(\ref{eq:two}). Similarly, we can also verify that any kind of
SOC could not give rise to spin polarization in the ssDNA.

According to the Landauer-B\"{u}ttiker (LB) formula, the current in
the $q$th lead (real or virtual) with spin $s$ can be written as
\cite{ds}: $ I_{q s}=(e^2/h) \sum_{m, s'}T_{qs,ms'} (V_{m}- V_{q})$,
where $V_{q}$ is the voltage in the $q$th lead and $T_{qs,ms'} ={\rm
Tr}[{\bf \Gamma}_{qs} {\bf G}^r {\bf \Gamma}_{ms'} {\bf G}^a]$ is
the transmission coefficient from the $m$th lead with spin $s'$ to
the $q$th lead with spin $s$. The Green function ${\bf G}^r=[{\bf
G}^a]^\dag =[E{\bf  I}- {\bf H}_{\rm DNA}-{\bf H}_{\rm so}-
\sum_{qs} {\bf \Sigma}_{q s } ^r ] ^{-1}$ and ${\bf \Gamma}_{q s}
=i[{\bf \Sigma}_{q s} ^r -{\bf \Sigma}_{qs} ^a]$, with $E$ the
incident electron energy (or the Fermi energy). ${\bf \Sigma}_{
qs}^r$ is the retarded self-energy due to the coupling to the $q$th
lead. For the real left/right lead, ${\bf \Sigma}_{{\rm L}/{\rm R}
s}=-i\Gamma_{{\rm L}/{\rm R}}/2 =-i\pi \rho_{{\rm L}/{\rm R}}
t^2_{{\rm L}/{\rm R}}$; while for the virtual leads, ${\bf
\Sigma}_{q s} ^r = -i \Gamma /2 =-i\pi\rho_d t^2_d$, with the
dephasing parameter $\Gamma$ and $\rho_{{\rm L}/{\rm R}/d}$ being
the density of state of the leads. Since the net currents through
the virtual leads are zero, their voltages can be calculated from
the LB formula by applying an external bias $V_{\rm b}$ between the
left and right leads with $V_{\rm L}=V_{\rm b}$ and $V_{\rm R}=0$.
Finally, the conductance for spin-up ($G_\uparrow $) and spin-down
($G_\downarrow $) electrons can be obtained $G_{s} = (e^2/h)
\sum_{m, s'}T_{{\rm R}s,ms'} V_{m}/V_{\rm b}$, and the spin
polarization is $P_{\rm s}=(G_\uparrow- G_\downarrow)/ (G_\uparrow+
G_\downarrow)$.

For the dsDNA, $\varepsilon_{jn}$ is set to $\varepsilon_ {1n}= 0$
and $\varepsilon_ {2n} = 0.3 $, $t_ {jn}$ is taken as $t_{1n}=0.12$
and $t_{2n}=-0.1$, and $\lambda_n =-0.3$. All these parameters are
extracted from first-principles calculations \cite{yyj,sk} and the
unit is eV. The helix angle and the twist angle are set to
$\theta=0.5$ rad and $\Delta\varphi ={\frac \pi 5}$, which are
typical values of B-form DNA. The SOC is estimated to $t_{\rm
so}=0.01$, which is an order of magnitude smaller than the
intrachain hopping integral. In fact, all the results are
qualitatively same even for smaller $t_{\rm so}$. For the real
leads, the parameters $\Gamma_{\rm L} = \Gamma_{\rm R} =1$ are
fixed. For the virtual leads, the dephasing strength is very small
with $\Gamma=0.005$, because the monolayers of the dsDNA are rigid
and the longest DNA in experiment is short \cite{gb}. For this value
of $\Gamma$, the phase coherence length is longer than the dsDNA and
the electron transport through the dsDNA will keep its phase
coherence \cite{xy}. However, the finite dephasing is indispensable.
The values of all above-mentioned parameters will be used throughout
the Letter except for specific indication in the figure.

\begin{figure}
\includegraphics[scale=1.]{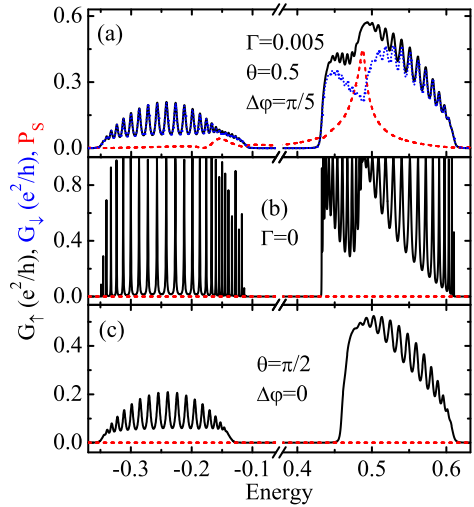}
\caption{\label{fig:two}(color online). (a) Energy-dependence of
conductance $G_{\uparrow}$ (solid line), $G_{\downarrow}$ (dotted
line), and spin polarization $P_{\rm s}$ (dashed line) for realistic
situation. (b) and (c) show $G_\uparrow$ and $P_{\rm s}$ in the
absence of the dephasing and of the helical symmetry, respectively.
Here N=30.}
\end{figure}

Figure~\ref{fig:two}(a) shows the conductances $G_{\uparrow /
\downarrow }$ and the corresponding spin polarization $P_{\rm s}$.
One notices two transmission bands---the highest occupied molecular
orbital (HOMO) and the lowest unoccupied molecular orbital
(LUMO)---in the energy spectrum, where several transmission peaks
are found for both spin-up and spin-down electrons (holes) due to
the coherence of the system. For the HOMO band, $G_{\uparrow}$ and
$G_{\downarrow}$ are almost identical, while for the LUMO band,
$G_{\uparrow}$ and $G_{\downarrow}$ are very different. A
bell-shaped configuration is observed in the curve of $P_{\rm s}$ vs
energy $E$, where the spin polarization of the dsDNA at $N=30$ can
achieve $0.45$, which is comparable with the experimental results
\cite{gb}.

To explore the physical scenario to high spin polarization observed
in the dsDNA, Figs.~\ref{fig:two}(b) and~\ref{fig:two}(c) plot the
conductance $G_{\uparrow}$ and $P_{\rm s}$ in the absence of the
dephasing ($\Gamma=0$) and of the helical symmetry ($\theta= \frac
\pi 2$), respectively. It clearly appears that the spin polarization
vanishes when $\Gamma=0$ or $\theta=\frac \pi 2$, although the
conductance is still very robust in both cases. When $\Gamma=0$, the
dsDNA decouples with the virtual leads and the charge transport
through the dsDNA is completely coherent. In this case, the SOC can
not generate any spin polarization due to the time-reversal
symmetry, the helical symmetry, and the phase-locking effect
\cite{sqf3}. A small dephasing is necessary for the existence of
finite spin polarization. Indeed, it is reasonable to assume a small
$\Gamma$ because the dephasing occurs inevitably in the experiment.
On the other hand, the spin polarization strongly depends on the DNA
helix. If there is no helix ($\theta=\frac \pi 2$), no spin
polarization could be observed ($P_{\rm s}=0$). If the right-handed
helical dsDNA is transformed into the left-handed one (Z-form DNA)
with $\theta \rightarrow \pi - \theta$, $P_{\rm s}(\pi-\theta)=- P_{
\rm s} (\theta)$ exactly.

\begin{figure}
\includegraphics[scale=0.77]{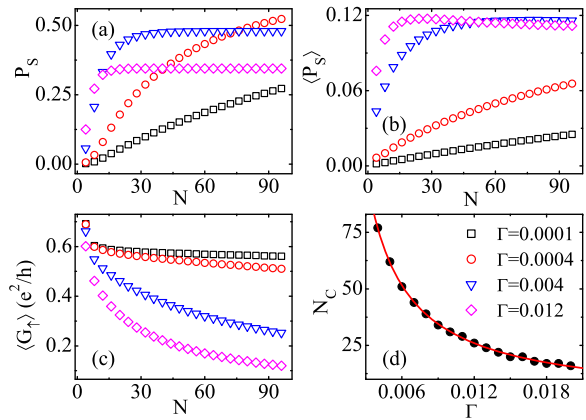}
\caption{\label{fig:three}(color online). Length-dependence (a) of
$P_{\rm s}$ at $E=0.488$, (b) of $\langle P_{\rm s}\rangle$, and (c)
of $\langle G_{\uparrow}\rangle$ for different values of the
dephasing parameter. (d) The critical length $N_c$ vs $\Gamma$. Here
$N_c$ is extracted from the curve of $\langle P_{\rm s} \rangle$-$N$
and the solid line is the fitting curve with $N_c \propto \Gamma ^ {
-1 } $. The legends in (d) are for panels (a), (b), and (c). }
\end{figure}

We then focus on the spin polarization $P_{\rm s}$ and the averaged
one $\langle P_{\rm s}\rangle$, where $\langle P_{\rm s} \rangle
\equiv ( \langle G_\uparrow \rangle-\langle G_\downarrow \rangle
)/(\langle G_\uparrow \rangle + \langle G_\downarrow \rangle)$ with
$\langle G_s \rangle$ averaged over the LUMO band.
Figures~\ref{fig:three}(a) and~\ref{fig:three}(b) show $P_{\rm s}$
at a fixed energy $E$ and $\langle P_{\rm s} \rangle$ vs length $N$,
respectively, for several values of the dephasing parameter. One
notices that $P_{\rm s}$ and $\langle P_{\rm s} \rangle$ are
enhanced by increasing $N$ at first and then saturate or slightly
decline after a critical length $N_c$. With increasing $\Gamma$,
$N_c$ shrinks monotonically [Fig.~\ref{fig:three}(d)]. The behavior
of $N_c$ vs $\Gamma$ can be fitted well by a simple function $N_c
\propto \Gamma ^ {-1}$. For relatively large $\Gamma$ (diamond and
triangle symbols), $P_{\rm s}$ and $\langle P_{\rm s} \rangle$
increase faster in the beginning and saturate at shorter length with
smaller values because the device is more open. While for smaller
$\Gamma$, $P_{\rm s}$ and $\langle P_ { \rm s} \rangle$ increase
slower with increasing $N$ in a wider range of $N$ and have larger
saturation values. Let's see $\Gamma=0.0004$ for instance (circle
symbols). $P_{\rm s}$ and $\langle P_{ \rm s} \rangle $ will keep
rising even for $N>100$, and $P_{\rm s}=0.34 $ at $N=40$ and $P_{\rm
s}= 0.5$ at $N=80$. These results are quantitatively consistent with
the experiment \cite{gb}. In fact, the dephasing has two effects:
(i) it promotes the openness of the two-terminal device and produces
the spin polarization \cite{sqf3}; (ii) it makes the charge lose its
phase and spin memories and then $P_{\rm s}$ is decreased by further
increasing $\Gamma$. Accordingly, for large $\Gamma$ with the phase
coherence length $L_{\phi}$ \cite{xy} shorter than the dsDNA length,
$P_{\rm s}$ will be quite small. $P_{\rm s}<0.05$ for $\Gamma=0.5$
and $P_{\rm s}\rightarrow 0$ if $\Gamma\rightarrow \infty$. Due to
the interplay between the above two effects, a small $\Gamma$,
ranging from 0.0002 to 0.01, where $L_{\phi}$ is much larger than
$100$, is optimal for large $P_{\rm s}$. In addition,
Fig.~\ref{eq:three}(c) shows the averaged conductance $\langle
G_{\uparrow} \rangle$ vs $N$. $\langle G_{\uparrow} \rangle$ is
declined by increasing $N$ or $\Gamma$, because large $N$ or
$\Gamma$ will enhance the scattering. However, $\langle G_{ \uparrow
} \rangle$ remains quite large for $N=100$ and $\Gamma =0.012 $.
Therefore, the dsDNA is a well spin filter due to the large $P_{\rm
s}$ and $\langle G_{\uparrow} \rangle$.

\begin{figure}
\includegraphics[scale=0.9]{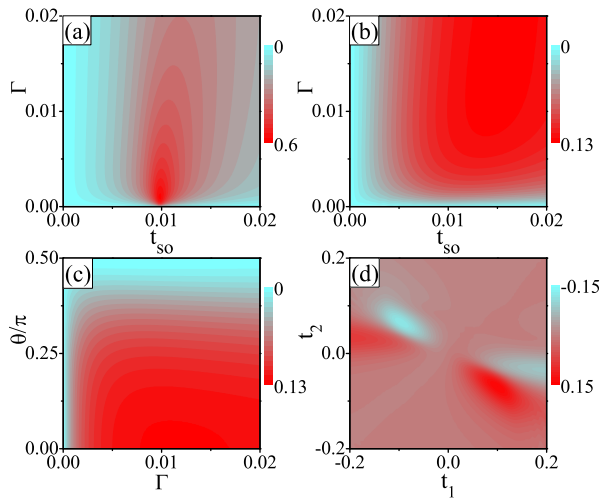}
\caption{\label{fig:four}(color online). (a) $P_{\rm s}$ vs the SOC
$t_{\rm so}$ and the dephasing $\Gamma$ at $E=0.488$ with $N=80$.
(b), (c), and (d) show $\langle P_{\rm s}\rangle $ with $N=30$ as
functions of $t_{\rm so}$ and $\Gamma$, of $\Gamma$ and ${\frac
\theta \pi}$, and of $t_1$ and $t_2$, respectively.}
\end{figure}

Let us further study the spin polarization by varying other model
parameters. Figures~\ref{fig:four}(a) and \ref{fig:four}(b) show
$P_{ \rm s}$ at $E=0.488$ with $N=80$ and $\langle P_{\rm s}\rangle$
with $N=30$, respectively, as functions of the SOC $t_{\rm so}$ and
the dephasing strength $\Gamma$. $P_{\rm s}$ and $\langle P_{\rm
s}\rangle $ are zero exactly when $\Gamma=0$ or $t_{\rm so}=0$. Of
course, $t_{\rm so}$ is a key factor for the spin polarization or
equivalently $t_{\rm so}$ is ``the driving force'' of $P_{\rm s}$.
If there is no SOC, no spin polarization would appear for whatever
other parameters are. In general, strong SOC usually lead to large
$\langle P_{\rm s} \rangle$ [Fig.4(b)]. However, for a fixed energy,
$P_{\rm s}$ will not increase monotonically with $t_{\rm so} $, as
seen in Fig.~\ref{fig:four}(a). A large $P_{\rm s}$ can be obtained
for long dsDNA even for quite small $t_{\rm so}$, because the spin
polarized electrons will accumulate gradually when electrons are
transmitting along the dsDNA. In addition, we observe a large area
with red color in Fig.~\ref{fig:four}(b), where $\langle P_{\rm s}
\rangle $ exceeds $0.1$ for short dsDNA. This implies that the dsDNA
would be an efficient spin filter in a wide parameters range.

Figure~\ref{fig:four}(c) plots the averaged spin polarization
$\langle P_{\rm s}\rangle$ vs $\Gamma$ and ${\frac\theta \pi}$ by
fixing the radius $R$ and the arc length $l_a$ to account for the
rigid sugar-phosphate backbones. The helix angle $\theta$ can be
changed by stretching the DNA molecule \cite{gj}. It is obvious that
$\langle P_{\rm s}\rangle$ is zero in the absence of the helical
symmetry ($\theta={\frac \pi 2}$) and $\langle P_{\rm s}\rangle$ is
increased monotonically by decreasing $\theta$. This indicates that
the helix of the dsDNA plays a vital role to the existence of the
spin polarization. Finally, we present the influence of the hopping
integrals $t_1$ and $t_2$ on the spin polarization, as illustrated
in Fig.~\ref{fig:four}(d). We can see that $\langle P_{\rm s}
\rangle $ is small when $t_1$ and $t_2$ have identical sign and
become large when $t_1$ and $t_2$ have opposite sign. Since the sign
of the hopping integral is sensitive to the type of neighboring
nucleobases \cite{sk} and to the twist angle $\Delta\varphi$
\cite{erg}, the spin polarization could be improved by synthesizing
specific DNA molecule and putting force along the helix axis of the
dsDNA.

In summary, we propose a model Hamiltonian to simulate the quantum
spin transport through the dsDNA. This two-terminal dsDNA-based
device would exhibit high spin polarization by considering the SOC,
the dephasing, and the helical symmetry, although no spin
polarization exists in the ssDNA. The spin polarization increases
with increasing the DNA length. Additionally, the spin polarization
could be improved by properly modifying the hopping integral and
decreasing the helix angle.

This work was supported by China-973 program and NSF-China under
Grants No. 10974236 and 10821403.

%\end{references}


\begin{thebibliography}{00}
\bibitem{note1} In other words, here we consider the SOC induced by the
boundary-confining potential. All the results are qualitatively same
if other kinds of SOC are considered.
\end{thebibliography}
\end{document}